\documentclass[3p,times,procedia]{elsarticle}
\usepackage{nupha_ecrc}
\usepackage{amsmath}

\volume{00}

\firstpage{1}

\journalname{Nuclear Physics A}

\runauth{L. Apolin\'{a}rio et al.}

\jid{nupha}

\jnltitlelogo{Nuclear Physics A}

\usepackage{amssymb}
\usepackage[figuresright]{rotating}

\begin{document}

\begin{frontmatter}



\dochead{XXV International Conference on Ultrarelativistic Nucleus-Nucleus Collisions}
%% e.g. \dochead{17th International Conference on Dynamical Processes in Excited States of Solids}

\title{In-medium jet evolution: interplay between broadening and decoherence effects}

\author[label1]{Liliana Apolin\'{a}rio}
\author[label2]{N\'{e}stor Armesto}
\author[label1,label3]{Guilherme Milhano}
\author[label2]{Carlos A. Salgado}

\address[label1]{CENTRA, Instituto Superior T\'{e}cnico, Universidade de Lisboa, Av. Rovisco Pais, 1049-001 Lisboa, Portugal}
\address[label2]{Departamento de F\'{i}sica de Part\'{i}culas and IGFAE, Universidade de Santiago de Compostela, 15706 Santiago de Compostela, Galicia-Spain}
\address[label3]{Physics Department, Theory Unit, CERN, CH-1211 Geneve 23, Switzerland}

\begin{abstract}
The description of the modifications of the coherence pattern in a parton shower, in the presence of a QGP, has been actively addressed in recent studies. Among the several achievements, finite energy corrections, transverse momentum broadening due to medium interactions and interference effects between successive emissions have been extensively improved as they seem to be essential features for a correct description of the results obtained in heavy-ion collisions. In this work, based on the insights of our previous work \cite{Apolinario:2014csa}, we explore the physical interplay between broadening and decoherence, by generalising previous studies of medium-modifications of the antenna spectrum \cite{MehtarTani:2010ma,CasalderreySolana:2011rz,MehtarTani:2012cy} - so far restricted to the case where transverse motion is neglected. The result allow us to identify two quantities controlling the decoherence of a medium modified shower that can be used as building blocks for a successful future generation of jet quenching Monte Carlo simulators: a generalisation of the $\Delta_{med}$ parameter of the works of \cite{MehtarTani:2010ma,MehtarTani:2012cy} - that controls the interplay between the transverse scale of the hard probe and the transverse resolution of the medium - and of the $\Delta_{coh}$ in \cite{Apolinario:2014csa} - that dictates the interferences between two emitters as a function of the transverse momentum broadening acquired by multiple scatterings with the medium.
\end{abstract}


\end{frontmatter}

% \linenumbers

\section{Introduction}
\label{label:intro}

\par The theoretical description of jet quenching phenomena - collection of medium-induced modifications that take place when hard probes propagate through a hot and dense medium, such as the Quark-Gluon Plasma (QGP) - has been continuously improved in these last years. Among them, there are several works that contributed to the complete description of single gluon emission processes \cite{Apolinario:2014csa,Apolinario:2012vy,Blaizot:2013vha,Blaizot:2012fh} and coherence effects within a QCD medium antenna \cite{MehtarTani:2010ma,CasalderreySolana:2011rz,MehtarTani:2012cy} - single gluon emission from a quark-antiquark pair,  resulting from the splitting of photon or a gluon, that propagates through QGP. Nonetheless, to describe simultaneously the experimental observations on jet energy loss \cite{Aad:2010bu}, its angular distribution \cite{PhysRevC.84.024906} and jet fragmentation functions \cite{Aad:2014wha,PhysRevC.90.024908}, none of these effects alone is sufficient \cite{Apolinario:2012cg,Zapp:2013vla,Casalderrey-Solana:2014bpa}. In this work, by choosing as a framework a non-eikonal QCD antenna (a QCD antenna that is allowed to have Brownian motion in the transverse plane), we are able to make a more complete description of jet quenching phenomena by including simultaneously coherence effects, energy loss and transverse momentum broadening. By exploring the findings from \cite{Apolinario:2014csa}, it is possible to obtain the gluon emission spectrum for this system.

\section{Particle propagation in a QCD medium: energy loss and transverse momentum broadening}
\label{label:sec1}

\par To study in-medium modifications to the vacuum parton branching one needs to first consider the propagation of a highly energetic particle through a hot and dense medium. Due to the fact that the energy scale of the probe is much larger than the one from the medium, it is usually assumed that the particle undergoes multiple soft interactions. Those induce a color field rotation and transverse momentum \emph{kicks} without degrading its longitudinal energy\footnote{Light-cone coordinates will be used throughout the manuscript whose relation with Minkowski coordinates is given by: $x_{\pm} = (x_0 \pm x_3)/\sqrt{2}$ and $\mathbf{x} = (x_1, x_2)$. Moreover, the gauge is fixed such that the color field component $A_+ = 0$.}. Within the path-integral formalism, the propagation of a hard probe with longitudinal momentum $p_+$ through a recoiless medium of length $L_+$, from $\mathbf{x_0}$ to $\mathbf{x}$, is given by a Green's function:
\begin{equation}
\label{eq:GreensFunction}
	\mathcal{G}(x_{0+}, \mathbf{x_{0}}; L_{+}, \mathbf{x}) = \int_{\mathbf{r}( x_{0+}) = \mathbf{x_0}}^{\mathbf{r}(L_+) = \mathbf{x}} \mathcal{D} \mathbf{r}(\xi) \exp \left\{ \frac{ip_+}{2} \int_{x_{0+}}^{L_+} d\xi \left( \frac{d\mathbf{r}}{d\xi} \right)^2 \right\} W(x_{0+}, L_+; \mathbf{r}(\xi)) \, ,
\end{equation}
where
\begin{equation}
\label{eq:WilsonLine}
	W(x_{0+}, L_+; \mathbf{r}(\xi)) = \mathcal{P} \exp \left\{ ig \int_{x_{0+}}^{L_+} d\xi A_-(\xi, \mathbf{r}(\xi)) \right\} \, ,
\end{equation}
is the Wilson line that follows an ordered path in the medium color fields $A_-(x_+, \mathbf{x})$ with which the particle interacts. To calculate the energy loss, one can consider the gluon bremsstrahlung process which accounts for the evaluation of the elementary gluon emission vertex inside a finite medium. The latter is seen as a collection of static scattering centres whose colour field configuration is frozen during the propagation time of the fast particle. As such, to account for the total cross-section, an average over the possible ensemble of medium configurations, schematically represented by $ \left\langle \cdots \right\rangle$, has to be performed. The latest results within this formalism \cite{Apolinario:2014csa,Blaizot:2013vha,Blaizot:2012fh} have shown that for an infinite medium, both final particles undergo independent broadening. However, when considering a finite medium, there is an additional contribution that is the probability that both final particles remain in a coherent state (as in vacuum). Such contribution appears when considering the medium average of the final particles at the level of the cross-section. Considering only the color structure propagators, $W$, of both final quark (q) and gluon (g), at transverse positions $\mathbf{x}_q$, $\mathbf{x}_g$ in the amplitude and $\bar{\mathbf{x}}_q$, $\bar{\mathbf{x}}_g$ in the complex conjugate amplitude, one gets\footnote{This result is for a large number of colours, $N$.}: 
\begin{equation}
\label{eq:quadrupole}
	\frac{1}{N} Tr \left\langle W^\dagger (\mathbf{x}_g) W  (\bar{\mathbf{x}}_g) W^\dagger (\bar{\mathbf{x}}_q) W (\mathbf{x}_q) \right\rangle \underset{N \rightarrow \infty}{=} \frac{1}{N^2}Tr \left\langle W^\dagger  (\mathbf{x}_g)  W  (\bar{\mathbf{x}}_g)  \right\rangle Tr \left\langle W^\dagger  (\bar{\mathbf{x}}_q)  W  (\bar{\mathbf{x}}_q)  \right\rangle \Delta_{coh} \, ,
\end{equation}
with
\begin{equation}
	\Delta_{coh} = 1 + \int_{x_{2+}}^{L_+} d\tau \hat{q}_F \left. (\mathbf{x}_q - \bar{\mathbf{x}}_q) \cdot (\bar{\mathbf{x}}_g - \mathbf{x}_g)\right|_{\tau} \exp \left\{ \hat{q}_F\int_{x_{2+}}^\tau d\xi \, (\mathbf{x}_q - \mathbf{x}_g) \cdot (\bar{\mathbf{x}}_g - \bar{\mathbf{x}}_q) \right\} \, ,
\end{equation}
where $\tau$ and $\xi$ are integration variables in the longitudinal direction and $\hat{q}_F$ the transport coefficient that characterises the transverse momentum squared, $\mu^2$, transferred from the medium to the projectile per mean free path, $\lambda$.
In the limit that the emission process is \textit{quasi-local} \cite{Blaizot:2013vha,Blaizot:2012fh}, $\Delta_{coh} \rightarrow 1$, both final particles decorrelate in colour and the parton shower factorizes completely loosing energy more efficiently.

\section{In-medium QCD antenna: eikonal vs non-eikonal}
\label{label:sec2}

\begin{figure}[!ht]
      \vspace{-6.5mm}
  \centering
   \includegraphics[width=0.9\textwidth]{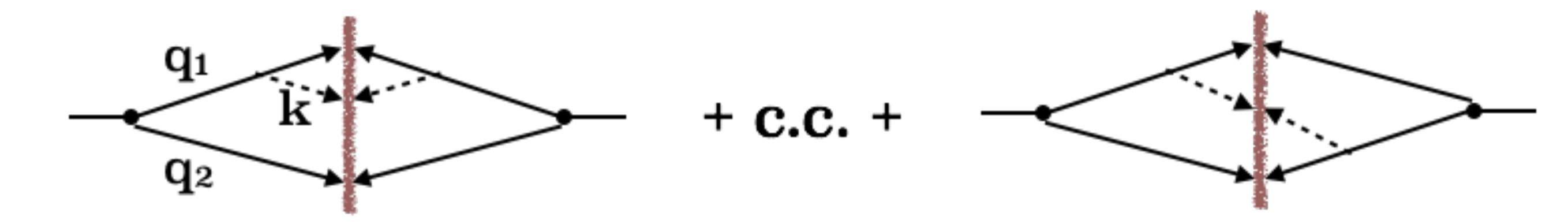} 
    \caption{Schematic diagram of the cross-section for a QCD antenna. The dashed line represents the gluon of momentum $k$ and the solid lines the quark, of momentum $q_1$, and anti-quark, of momentum $q_2$. The red thick line separates the amplitude (on the left) from the conjugate amplitude (on the right).}
    \label{fig:Antenna}
      \vspace{-2.5mm}
\end{figure}

As seen from the previous section, the $\Delta_{coh}$ parameter already contains some of the coherence effects that may arise when two emitters of the same parton shower are allowed to emit. Nonetheless, to capture the full coherence physics, the ideal laboratory is a QCD antenna (diagrams for the total cross-section explicited in figure \ref{fig:Antenna}). In vacuum, this picture originates the angular ordering property that is the basis of Monte Carlo generators in proton-proton collisions. For a singlet antenna (an antenna initiated by a photon), it follows that the gluon energy spectrum, $dI/d\Omega_k$, is given by:
\begin{equation}
\label{eq:antenna_vac}
	\frac{dI}{d\Omega_k} = \alpha_s C_F (R_q + R_{\bar{q}} - 2 J) = \alpha_s C_F \frac{q_1 \cdot q_2} {(k\cdot q_1) (k\cdot q_2)} \equiv R_{coh} \, ,
\end{equation}
with $C_F = (N^2-1)/(2N)$, $N = 3$, $d\Omega_k = 1/(2k_+) 1/(2\pi)^3 dk_+ d\mathbf{k}$ the gluon phase space, $R_q$, $R_{\bar{q}}$ the contributions from both direct terms (gluon emission from quark and antiquark - first two diagrammatic terms of figure \ref{fig:Antenna}), and $-2 J$ the contribution from the interference term (last diagram of figure \ref{fig:Antenna}). 
%They are given by:
%\begin{equation}
%	R_{q} \propto \frac{q_{1+}}{(q_1 \cdot k)} \, ,\text{ first diagram of figure \ref{fig:Antenna}} \, ,
%\end{equation}
%\begin{equation}
%	R_{\bar{q}}\propto \frac{q_{2+}}{(q_2 \cdot k)} \, ,\text{ complex conjugate (c.c.) of first diagram of figure \ref{fig:Antenna}} \, ,
%\end{equation}
%\begin{equation}
%	2 J \propto \left[ \frac{q_{1+}}{(q_1 \cdot k)} + \frac{q_{2+}}{(q_2 \cdot k)} - \frac{ k_+ (q_1 \cdot q_2) }{ (k\cdot q_1) (k \cdot q_2) } \right] \, , \text{ third diagram of figure \ref{fig:Antenna}} \, .
%\end{equation}

\par In the presence of a medium, both quark and antiquark will undergo multiple scatterings with the medium constituents. Assuming an eikonal approximation, i.e., considering only Wilson lines (eq. \eqref{eq:WilsonLine}) as propagators so the only effect of the medium is a color rotation, the interference term\footnote{In the soft limit approximation. For the general case, see \cite{MehtarTani:2010ma,MehtarTani:2012cy}.}, will have a non-trivial colour structure:
\begin{equation}
\label{eq:Interference}
	-2 J \frac{1}{N} Tr \left\langle T^a W (\mathbf{x}_q) W^\dagger (\mathbf{x}_{\bar{q}}) T^b W (\mathbf{x}_{\bar{q}}) W^\dagger (\mathbf{x}_q) \right\rangle = -2 J (1- \Delta_{med}) \, ,
\end{equation}
where $\mathbf{x}_q$ and $\mathbf{x}_{\bar{q}}$ are the transverse coordinates of the quark and antiquark (notice that in the eikonal limit, the transverse coordinates of both amplitude and conjugate amplitude are fixed and are the same). $T^a$ are the colour matrices and the $\Delta_{med}$ parameter \cite{MehtarTani:2010ma,MehtarTani:2012cy}:
\begin{equation}
	\Delta_{med} \approx 1 - \exp \left\{ - \frac{1}{12} \hat{q} \, \theta_{q\bar{q}} L_+ \right\} \, ,
\end{equation}
being $L_+$ the medium length and $\theta_{q\bar{q}}$ the angle between the quark-antiquark. For a very dense medium, $\Delta_{med}$ approaches the unit and there is complete decoherence between the two emitters (the two particle propagate independently), while in the opposite limit one recovers the vacuum coherent spectrum. As such, the antenna radiation spectrum takes the form
\begin{equation}
	\frac{dI}{d\Omega_k} \propto R_{coh} + 2 J \Delta_{med} \, .
\end{equation}

\par In order to describe a non-eikonal QCD antenna to account for transverse Brownian motion of both quark and antiquark, Green's functions (eq. \eqref{eq:GreensFunction}) should be used instead of Wilson lines. The color structure of both direct terms is now also modified by the same structure as in eq. \eqref{eq:quadrupole}, while the interference term is just a generalisation of the previous $\Delta_{med} \rightarrow \overline{\Delta_{med}}$ beyond the eikonal limit, i.e., considering different transverse coordinates in both amplitude and conjugate amplitude in eq. \eqref{eq:Interference}. Joining everything, one can write schematically that:
\begin{equation}
\begin{split}
	\frac{dI}{d\Omega_k d\Omega_1 d\Omega_2} & \propto (R_q + R_{\bar{q}}) \frac{1}{N^2} Tr \left\langle \mathcal{G}^\dagger_2 \mathcal{G}_{\bar{2}} \right\rangle Tr \left\langle \mathcal{G}_1 \mathcal{G}^\dagger_{\bar{1}} \right\rangle \Delta_{coh} - 2 J (1 - \overline{\Delta_{med}}) \\
	&= (R_q + R_{\bar{q}}) \Delta_{coh}^\prime - 2 J (1 - \overline{\Delta_{med}}) \, .
\end{split}
\end{equation}
where it was abbreviated $\mathcal{G}_{i (\bar{i})} = \mathcal{G} (x_{0+}, \mathbf{x}_{0\perp} (\bar{\mathbf{x}}_{0\perp}); L_+, \mathbf{x}_i (\bar{\mathbf{x}}_i))$ and the subscript 1(2) refer to the (anti)quark. In this form, the physical interpretation relies on a diagrammatic separation between direct terms and interference term. The former behave as two independent sources for gluon emission as in section \ref{label:sec1}. The latter continues to have the separation between the transverse resolution of the antenna $r_\perp = \theta_{q\bar{q}} L$, and the medium scale $Q_s^2 = \hat{q} L$, both with an additional dependence on the medium path. As such, the same physical picture holds, but now including energy loss and broadening effects. It is also possible to write the above expression by separating in \textit{coherent} and \textit{decoherent} part of the spectrum by identifying $\Delta_{coh}^\prime - \Delta_{med}^\prime = 1 - \overline{\Delta_{med}}$. In this case:
\begin{equation}
	\frac{dI}{d\Omega_k d\Omega_1 d\Omega_2} =  R_{coh} \Delta_{coh}^\prime + 2 J \Delta_{med}^\prime \, ,
\end{equation}
resulting in a non-trivial interplay between broadening, energy loss and coherence effects as we have a complete separation of angular ordering regimes, each enhanced/suppressed by coherence/decoherence effects.

\section{Conclusions}
\label{label:concl}

\par Color coherence effects have been studied for in-medium radiation and the interplay between coherence/decoherence and decoherence/broadening brought to theoretical control. The new calculation of a non-eikonal antenna in the soft limit generalises previous results. In addition, we found corrections due to the inclusion of Brownian motion and finite medium effects that seem to enhance even more the presence of hard vacuum-like subjets. As such, this work supports the picture of a medium-modified jet whose hard and central core is unmodified by the medium, a conclusion that is consistent with the experimental observations \cite{CasalderreySolana:2012ef}.
\newline
\newline
\textbf{Acknowledgements}: This work was supported by: Funda\c{c}\~{a}o para a
Ci\^{e}ncia e a Tecnologia of Portugal under project SFRH/ BPD/103196/2014 and CERN/FIS-NUC/0049/2015; the European Research Council grant HotLHC ERC-2011-StG-279579; Ministerio de Ciencia e Innovaci\'{o}n of Spain under project FPA2014-58293-C2-1-P; Xunta de Galicia (Conseller\'{i}a de Educaci\'{o}n) - the group is part of the Strategic Unit AGRUP2015/11.
\vspace{-3.5mm}



%% References


\bibliographystyle{elsarticle-num}
\bibliography{Bibliography}






% \end{thebibliography}

\end{document}